\pdfoutput=1
\documentclass[11pt]{article}

\usepackage[]{EMNLP2023}

\usepackage{times}
\usepackage{latexsym}
\usepackage{graphicx}
\usepackage{subfigure}
\usepackage{float}
\usepackage{caption}
\usepackage{amsmath}
\usepackage{amssymb}

\usepackage[T1]{fontenc}
\usepackage[utf8]{inputenc}

\usepackage{microtype}

\usepackage{inconsolata}

\title{Gated Deeper Models are Effective Factor Learners}

\author{Jingjing Guo \\
   Viterbi School of Engineering \\
  University of Southern California \\
  }

\begin{document}
\maketitle
\begin{abstract}
Precisely forecasting the excess returns of an asset (e.g., Tesla stock) is beneficial to all investors. However, the unpredictability of market dynamics, influenced by human behaviors, makes this a challenging task. In prior research, researcher have manually crafted among of factors as signals to guide their investing process. In contrast, this paper view this problem in a different perspective that we align deep learning model to combine those human designed factors to predict the trend of excess returns. To this end, we present a 5-layer deep neural network that generates more meaningful factors in a 2048-dimensional space. Modern network design techniques are utilized to enhance robustness training and reduce overfitting. Additionally, we propose a gated network that dynamically filters out noise-learned features, resulting in improved performance. We evaluate our model over 2,000 stocks from the China market with their recent three years records. The experimental results show that the proposed gated activation layer and the deep neural network could effectively overcome the problem. Specifically, the proposed gated activation layer and deep neural network contribute to the superior performance of our model. In summary, the proposed model exhibits promising results and could potentially benefit investors seeking to optimize their investment strategies.
\end{abstract}

\section{Introduction}
\noindent Deep learning has emerged as a powerful technique in the interdisciplinary field of computational finance, particularly in the area of quantitative factor mining. The promising capability of deep learning methods in analyzing large, complex, and high-dimensional datasets has enabled researchers to develop more accurate models for predicting financial market movements. Traditional methods, such as linear regression and time series analysis, have shown limitations in handling the intricate nature of financial data, thus leading to an increased interest in applying advanced deep learning techniques for better predictions and decision-making. 

In the past decade, a rapid growing research have been conducted to explore adopting deep learning methods in computational finance. For instance, \citep{xiong2015deep} employed an LSTM neural network to model the volatility of the S\&P 500 index, incorporating Google domestic trends as proxies for public sentiment and macroeconomic factors. Their work showed applicability to speech time series data. \cite{singh2017stock} applied deep learning to stock prediction, evaluating its performance on Google stock price multimedia data sourced from NASDAQ. \cite{gao2016deep} aimed to achieve high precision in stock market forecasting using deep learning methods, with a focus on Stock Technical Indicators (STIs) \cite{agrawal2019stock}. \cite{lin2018application} utilized an LSTM model to learn and forecast the stock market valuation indicator, price-earnings ratio (P/E ratio). \cite{jiang2021applications} provided a comprehensive review of recent progress in deep learning applications for stock market prediction. Researchers have been exploring various machine learning tools with the objective of building effective prediction models.

In this paper, our goal is to construct a novel neural network model that combines Feedforward networks and Gated activation layers to predict whether a stock's future excess returns will be positive. Excess returns refer to the portion of a stock's performance that exceeds the market benchmark. Our proposed model addresses some limitations of previous models by incorporating advanced deep learning techniques and leveraging cutting-edge computational resources. The significance of this model lies in its potential to improve investment decision-making, enhance portfolio optimization, and facilitate risk management, ultimately contributing to the advancement of the computational finance industry.
\section{Preliminary}
%In this study~\cite{he2016deep,monti2018avoiding}, we focusing on solving the text classification problem by using data generation under the zero-shot setting.
%Specifically, given a downstream task scenarios $\mathcal{X}$ and a semantic label set $\mathcal{Y}$, the text classification problem aims at obtaining a classifier $f:\mathcal{X}\rightarrow\mathcal{Y}$ that maps any input $\tilde{x}=[w_1, ..., w_n, ..., w_N]\in\mathcal{X}$ into a semantic label $y\in\mathcal{Y}$, where each word $w_n$ comes from a pre-defined vocabulary set $\mathcal{V}$. 
%Traditionally, we train the classifier $f$ on a training dataset $\mathcal{D}_\mathrm{train}=\{(x, y)\}$, where each input  $x~\mathcal{X}$ is sampled from the real world scenarios, and each label $y$ is annotated by a human expert. 
%However, under the zero-shot setting, we cannot obtain the training dataset, i.e. $\mathcal{D}=\emptyset$. 
%To obtain a synthesis dataset $\mathcal{D}=\{(x, y)\}$ for training the classifier $f$ under the zero-shot setting (no sample is available), we generate some samples $x\sim g(\cdot|y)$ by using a language model $g$ for each label $y\in\mathcal{Y}$. 
%A better synthesis dataset should help the classifier $f$ performs better on the real cases $(\tilde{x}, y)$.

\noindent In this paper, we mainly focus on identify the effective Alphas that can determine whether the future excess returns of stocks will be positive, that is, whether the future price changes of stocks can outperform the market average. Comparing with forecasting the future returns of stocks, forecasting the direction of excess returns of stocks is more straightforward and could eliminate noise to faithfully evaluate the quality of factors. 
%can eliminate a lot of extraneous noise.
%thereby enhancing the accuracy of our predictions.

\subsection{Quantitative trading alphas}
The concept of Alpha ($\alpha$) was originally derived from the capital asset pricing model (CAPM) \cite{sharpe1964capital}, %An alpha refers to a model used to try to forecast the prices, or returns, of financial instruments relative to a benchmark.
where it considers a linear relationship between the return of the asset $i$ and that of the market portfolio $m$ at each time step $t$:
\begin{equation}
r_{i,t}- r_{f,t} = \alpha + \beta(r_{m,t} - r_{f,t}) + \varepsilon_i, 
\label{CAPM}
\end{equation}
where $r_{i,t}$ is the return of the asset $i$, $r_{f,t}$ is the risk-free rate, $r_{m,t} - r_{f,t}$ refers to the expected return of the market, and $\varepsilon_i$ is the error term which can not be captured by this linear model. 
Then the excess return $e_{i,t}$ is defined as the left-hand side of the equation of {CAPM}:
\begin{equation}
e_{i,t} = r_i - r_{f}.
\end{equation}

\noindent According to the above formula, portfolio returns can be divided into two parts: $\beta$ return, which is the return compensation for bearing systematic (market) risk calculated by CAPM, and $\alpha$ return, which is the difference in returns compared to the market benchmark and is not affected by market movements. Therefore, $\alpha$ is considered a better measure of a stock's true value and is most commonly used in multi-factor stock selection models.
Here, the return made by the alphas $\alpha$ and system $\beta$ are respectively defined as:
\begin{equation}
r_{\alpha, t} = (r_{i,t} - r_{f,t}) - \beta(r_{m,t} - r_{ft}),
\end{equation}
and 
\begin{equation}
r_{\beta, t} = \beta(r_{m,t} - r_{f,t}).
\end{equation}
In this study, we focusing on constructing better factors to achieve higher alpha return $r_{\alpha,t}$.

\subsection{Multi-factor stock selection models}
A multi-factor stock selection model is a quantitative investment strategy that uses a combination of different financial factors to predict the future performance of stocks. Alpha factor is a critical component of multi-factor models, representing the excess return generated by a stock beyond its exposure to market risk. These alpha factors can be derived from a range of metrics, including fundamental indicators like earnings, revenue, and book value, as well as technical indicators like moving averages and momentum.%\\ 
%As a result, alpha factors are an essential tool for investors seeking to generate higher returns than the market average.

In this study, we aim to generate more factors that could effectively capture excess return of stocks $e_{i,t}$ for each stock $i$ and time step $t$. 
To achieve this goal, we first introduce 101-Alpha factors~\cite{Kakushadze2015101FA}, which are dedicated designed by combining most popular price-volume-based signals (e.g., close-to-close returns, open price, close price, high price, low price, volume and vwap) with some financial formulations. 
The detail definitions of each factor in 101-Alpha is presented in the original article.
%The main idea of our model is to input 101 Alpha factors into our designed neural network model to get more factors which can captures excess return of stocks better. The alpha factors included are mostly "price-volume" (daily close-to-close returns, open, close, high, low, volume and vwap) based and the formulaic expressions of these alpha factors are shown in the article "101 Formulaic Alphas".%are given in Appendix A.

%%%%%%%%%%%%%%%%%%%%%%%%%%%%%%%%%%%%%%%%%%%%%%%%%%%%%%%%%%%%%%%%%%%%%%%%%%%%%%%%%%%%%%%%%%%%%%%%%%%%%%%%%%%%%%%
\section{Method} 

\subsection{Architecture} 
\begin{figure}[htbp]
    \centering
    \includegraphics[scale = 0.8]{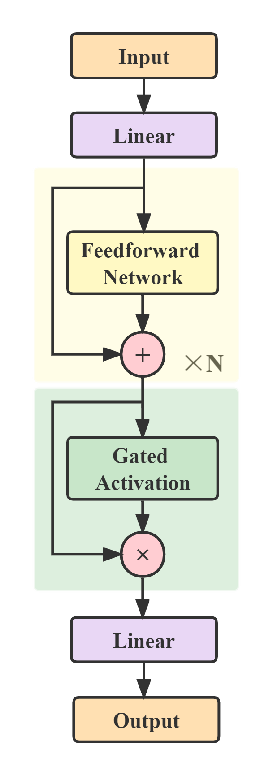}
    \caption{Structure of GatedDeep MLP model.}
    \label{total}
    \end{figure}

In this section, we present of our design for predicting the trend of future excess return of assets by aligning Alpha 101 factors. 
The model is primarily composes of two main modules, namely the Feedforward Networks and the Gated Activation Layer, where the former one is responsible to generating more meaningful in a higher dimensional space, while the later one dynamically select the new generated features. 
We also introduce batch normalization, skip connections, and dropout strategies into our model design to alleviate the degradation and overfitting problems for a robust training process. 
Figure \ref{total} shows the overall architecture of our proposed model.

\begin{figure}[h]
%\vspace{-0.3cm}
\centering
\subfigure[Feed Forward]
{\includegraphics[scale=0.145]{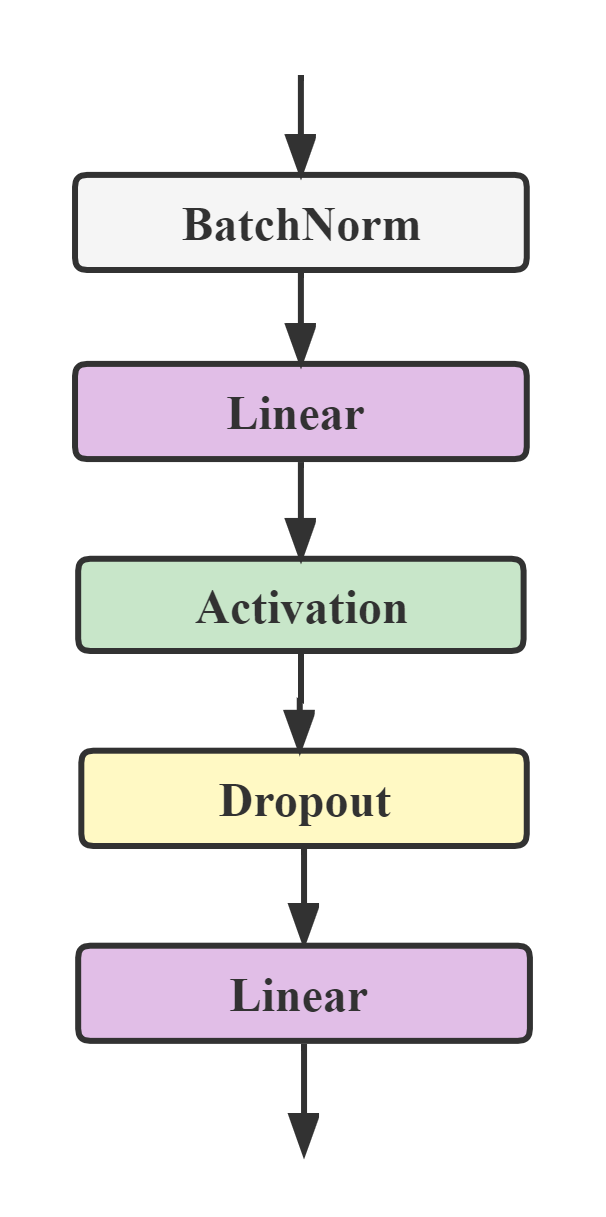}}\label{va}
\hspace{+0.4cm}
\subfigure[Gated Activation]
{\includegraphics[scale=0.145]{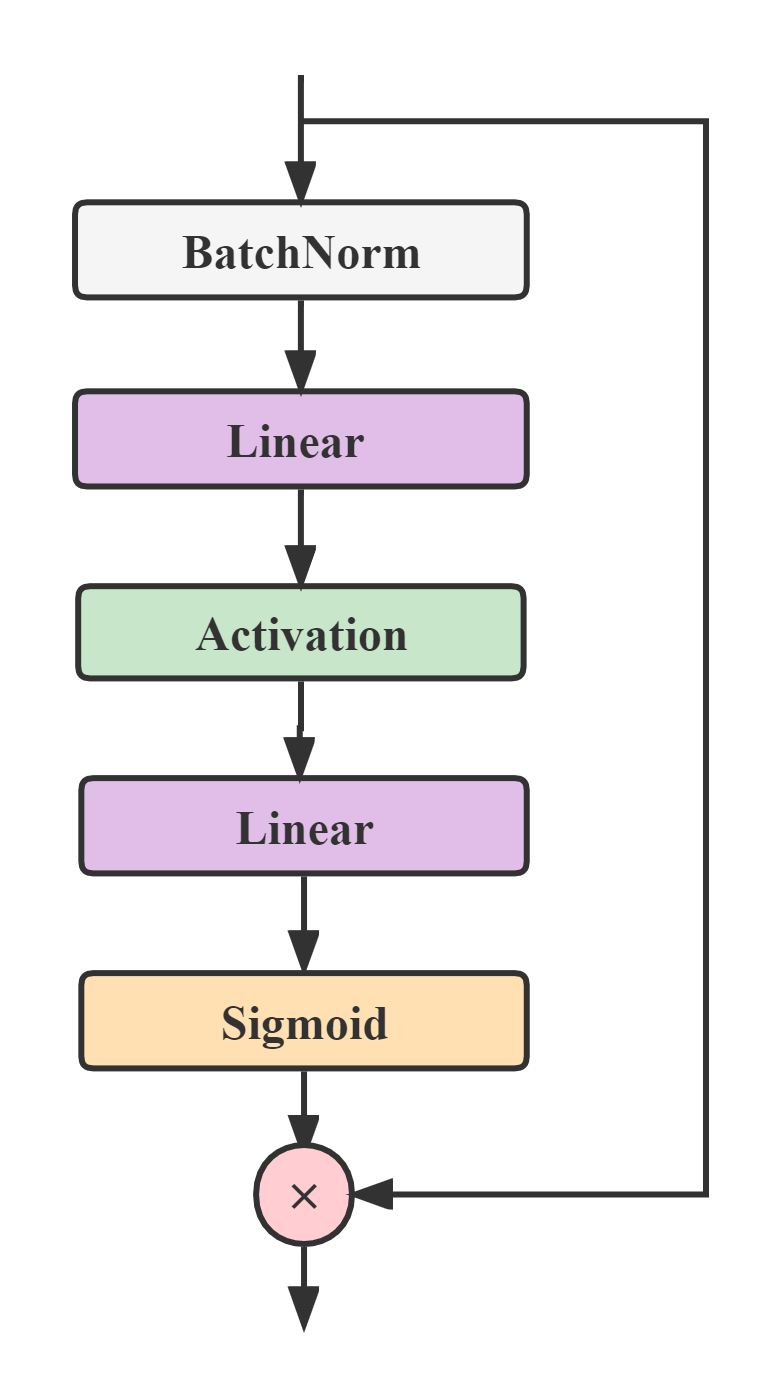}\label{ga}}
\vspace{-0.3cm}
\caption{Details of model architecturel}
\label{Fig:2}
\vspace{-0.6cm}
\end{figure}

\subsection{Feedforward Network}
%As is shown in Figure 2, the Feedforward Network is constructed by BatchNorm, Linear,
%Inspired by VGG nets, our Feedforward Network (FFN) (as shown in Figure 2) mainly consists of a batch normalization layer, a linear transformation layer, an activation function, a dropout layer, and another linear transformation layer. The first linear transformation layer expands the d-dimensional factors to d4-dimensional factors, while the second linear transformation layer compresses the d4-dimensional factors back to d-dimensional factors. Notably, the FFN architecture details are not crucial for our model. The total number of layers and weights may vary depending on the specific implementation.
The naive Feedforward Network module includes two linear transformation layers $\mathrm{linear}: \mathbb{R}^d\rightarrow\mathbb{R}^{m\times d}$ and one non-linear (activation) layer $f_\mathrm{activate}: \mathbb{R}^d\rightarrow\mathbb{R}^d$. 
Specifically, we first use the first linear layer transform the input embedding $x\in\mathbb{R}^d$ into a ($m\times d$)-dimensional representation $z$, then an activation layer is applied to operate the embedding $z$. Here, the activation function is $\mathrm{ReLU}(x) = max(0, x)$. Finally, the second linear layer is applied to map the high dimensional representation $z$ back into the original space $x'$. 
We expect the linear layers could learn an appropriate mapping between the input and output dimensions, thus enabling the embedding layers to capture and represent the underlying patterns and relationships within the input data. 

%\begin{figure}[htbp]
%    \centering
%    \includegraphics[scale = 0.12]{feedforward network.png}
%    \caption{Feedforward Network}
%    \end{figure}
%一种水平图，一种垂直图，取其一

%The phenomenon of degradation in deep neutral networks has been observed in many applications, including image classification and language modeling.It occurs when gradients become too small to effectively update the weights of the neural network during training, particularly in the earlier layers of the network. This can result in the network being unable to learn useful features, and can lead to the degradation of performance as the network becomes deeper.As the depth of a neural network is increased, accuracy reaches a saturation point and then declines rapidly. 
\noindent Generally, stacking feedforward modules could increase the number of model capacity, enabling it to learn more complex feature representations and improve its performance. However, in some cases, adding more layers to a neural network may cause its performance to degrade, resulting in decreased accuracy as the depth of the network increases. This phenomenon is commonly referred to as the \emph{degradation} problem, where the increased depth of the network causes the gradients to become smaller and prevents the model from properly learning and updating its weights. In this paper, we address this problem mainly by using batch normalization and residual connection. 
The residual connection is applied over each Feedforward module. That is, the output of each layer is $X_{i+1} = X_{i}+ f({X_i})$, where $f(X_{i})$ denotes a Feedforward layer. To facilitate these residual connections, all embedding layers produce outputs of dimension $d$. Additionally, batch normalization is introduced at the beginning of each feedforward module so that the strong dependency between two layers are decomposed. Finally, we adopt the dropout strategy over the outputs of the activation layer to encourage the second linear layer to learn more diverse and robust patterns from the high dimensional space. The strucuture of our feedforward network is presented in Figure 2(a).

%让我想想这句话之后再加点啥
\paragraph{Batch Normalization (BN).}
The basic idea behind batch normalization is to normalize the input data for each mini-batch at each layer of the neural network, such that the mean is close to 0 and the variance is close to 1.
The BN layer first determines the mean $\mu$ and the variance $\sigma^2$ of the activation values across the batch:
\begin{equation}
\begin{aligned}
\mu = \frac{1}{n} \sum_{i} Z_{i}, \,\,
\sigma^2 = \frac{1}{n} \sum_{i} (Z_{i} - \mu)^2.
\end{aligned}
\end{equation}
Then the activation vector $Z_{i}$ is normalized so that each neuron's output follows a standard normal distribution across the mini-batch. The constant epsilon $\varepsilon$ is used for numerical stability purposes:
\begin{equation}
Z_{i}^{'} = \frac{Z_{i}-\mu}{\sqrt{\sigma^2-\varepsilon}}.
\end{equation}
Finally, the layer's output $Z_{i}^{''}$ is calculated by applying a linear transformation with two trainable parameters: $\gamma$ and $\beta$:
\begin{equation}
Z_{i}^{''} = \gamma * Z_{i}^{'} + \beta,
\end{equation}
where $\gamma$ is used to adjust the standard deviation and $\beta$ is used to adjust the center of the distribution. 
%可以放入PPT演讲稿：If we increase $\beta$, the center of the distribution will be shifted to the right, while decreasing $\beta$ will shift it to the left.
 
\begin{figure}[]
    \centering
    \includegraphics[scale = 0.15]{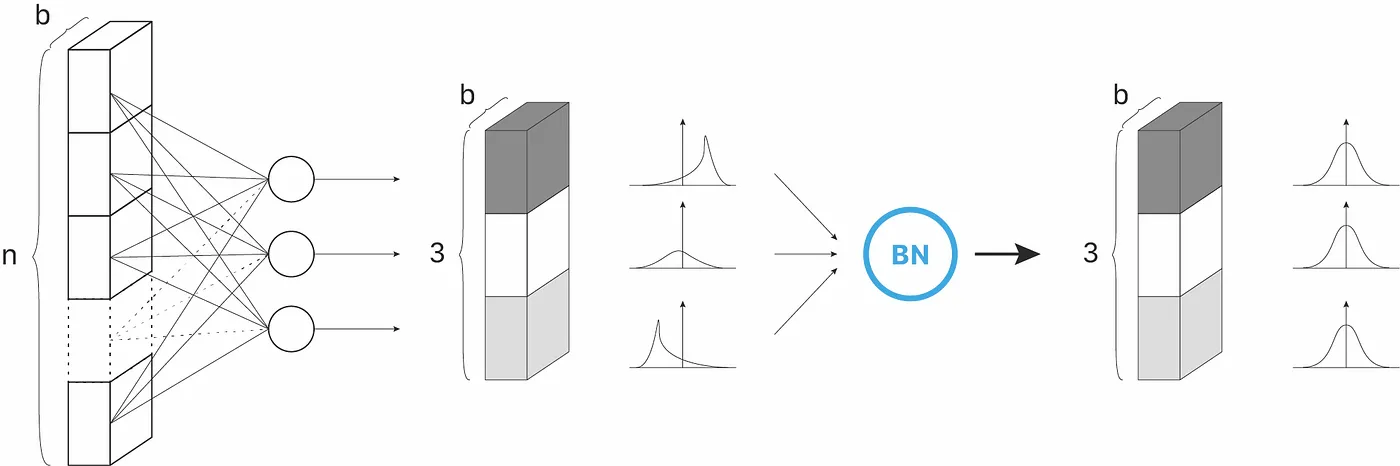}
    \caption{Batch Normalization | The outputs of each neuron will be first mapped to a distribution with zero mean and one standard deviation. Then they are later rescaled and shifted to a new distribution with mean and standard deviation dynamically learned during training. (Johann Huber, 2020)}
    \end{figure}
% Batch normalization的示意图：https://towardsdatascience.com/batch-normalization-in-3-levels-of-understanding-14c2da90a338#ad2e
% Batch Normalization: Accelerating Deep Network Training by  Reducing Internal Covariate Shift
%https://arxiv.org/pdf/1502.03167.pdf

\begin{figure}[htbp]
    \centering
    \includegraphics[scale = 0.4]{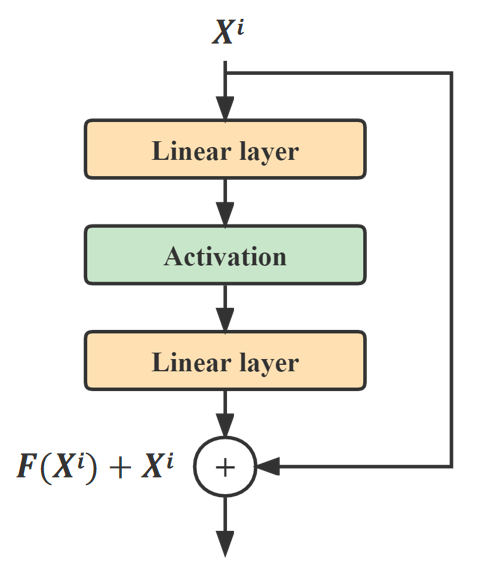}
    \caption{Residual learning~\cite{He2015DeepRL}.}
    \label{RC}
    \end{figure}

\paragraph{Residual Connection.} If we represent the formally underlying relationship between the input ($\mathcal{X}$) and output ($\mathcal{Z}$) as $h_j:\mathcal{X}\rightarrow\mathcal{Z}$ for each neuron $j$, we can train stacked nonlinear layers to fit an alternative mapping of $f(X):=h(X)-X$. This allows us to transform the original mapping into $f(X)+X$. The rationale behind this approach is that it is generally easier to optimize the residual mapping than the original, as it involves fitting a smaller and simpler function~\cite{c2deabe31641461c9c59d8362612c543, 627e99bb76a5475497f3febd817c85ae, monti2018avoiding}.\\
\noindent When the dimensions of $\mathcal{X}$ and $\mathcal{Z}$ are unequal, we will perform a linear projection $\mathcal{W}$ by the shortcut connections to match the dimensions:
\begin{equation}
Z = f(X)+ \mathbf{W}X,
\end{equation}
where $X$ and $Z$ are the input and output vectors of the layers; $f(x)$ its the residual mapping to be learned and $\mathbf{W}$ is a linear projection.\\
\noindent Based on the FeedForward network, we insert shortcut connections which turn the network into its counterpart residual version and presented in Figure~\ref{RC}. Through the linear projection in the above network, the dimensions of the outputs of layer 1 and layer 2 are the same and the identity shortcuts can be directly used. 
%Avoiding Degradation in Deep Feed-Forward Networks by Phasing Out Skip-Connections
%Deep Residual Learning for Image Recognition
%we add skip connections that allow for the passing of information across multiple layers. These connections add the output of one layer to the input of another.

%By providing a direct path for information to flow through the network, residual connections make it easier to optimize deep networks and improve overall performance.

\paragraph{Dropout.}
%Overfitting is a common problem in machine learning where a model becomes too complex and starts to fit the training data too closely, leading to poor performance on new testing data.
Dropout is a kind of computationally cheap and remarkably effective regularization method for reducing the overfitting problem and improving the generalization error in deep neural networks. It randomly sets a fraction of the activations in the neural network to zero during training, which forces the network to learn more robust features and reduces its reliance on any one activation.\cite{article}
\begin{figure}[htbp]
%\vspace{-0.3cm}
\centering
\subfigure[Standard Neural Net]
{\includegraphics[scale=0.22]{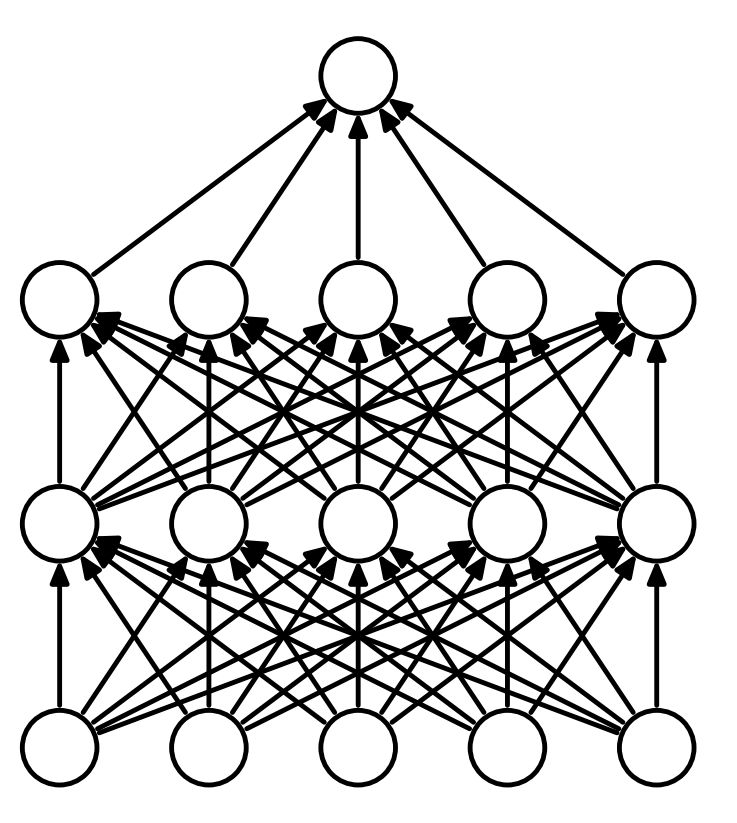}}\label{val}
\hspace{+0.4cm}
\subfigure[After applying dropout]
{\includegraphics[scale=0.22]{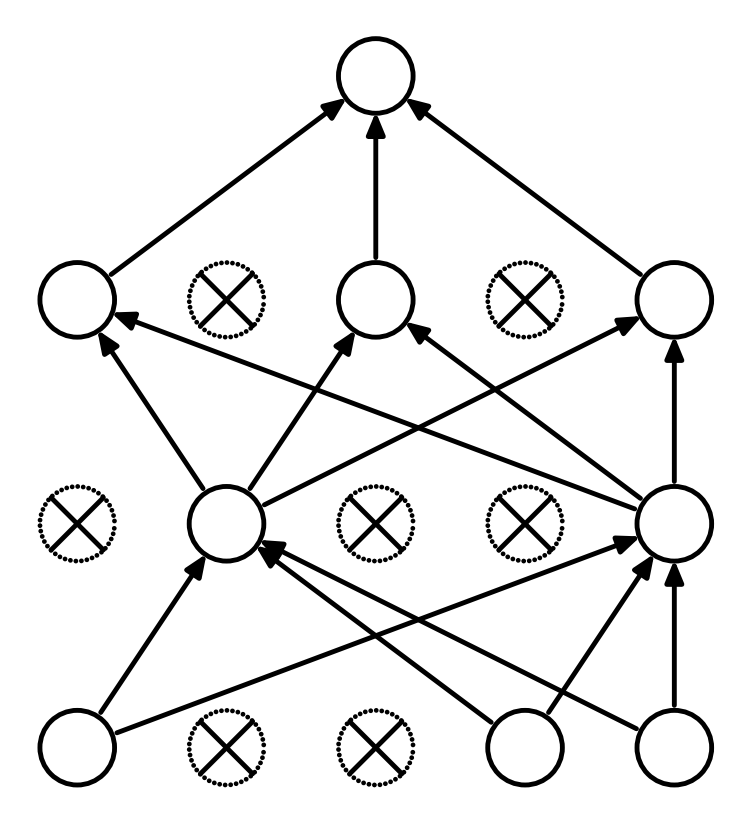}\label{vam}}
\vspace{-0.3cm}
\caption{Dropout strategy~\cite{article}.}
\label{Fig:2}
\vspace{-0.6cm}
\end{figure}

% Dropout: A Simple Way to Prevent Neural Networks from Overfitting
%https://dl.acm.org/doi/pdf/10.5555/2627435.2670313 

\subsection{Gated Activation Layer}
%d: 隐含层有多少个神经元 ~ d=512

An essential aspect of constructing the Gated Activation Layer is its ability to perform denoising, which facilitates faster and more accurate model inference and improves the model's robustness. As is shown in Figure~\ref{ga}, the Gated Activation Layer employs two linear transformations in a manner opposite to the Feedforward approach. The first linear transformation compresses the $d$-dimensional data to $d/k$ dimensions, and the second one rescales it back to $d$ dimensions. After these transformations, we use the sigmoid function to extract effective factors from the reconstructed $d$ dimensions, while the information of ineffective factors is compressed during the compression process. Finally, we multiply the sigmoid-filtered output on the original layer input to eliminate noise, resulting in a refined output.

\section{Experiments}
\subsection{Datasets}
\noindent The main source of our data includes Wind Information Inc.(WIND) and China Stock Market \& Accounting Research Database (CSMAR). The period of our data is from January 1, 2020, through March 1, 2023. Considering that institutions, organizations and individuals in China mainly trade in A-share market, we only select China A-shares that trade on the Shanghai Stock Exchange (SSE) in this paper. \\
\noindent Specifically, the stock data we need including daily technical data like open (opening price), close (closing price), high (daily highest price), low (dailylowest price), returns (daily close-to-close returns), volume(daily trading volume), vwap (daily volume-weighted average price), cap (market cap) and other fundamental data like cap(market cap) and ind (a generic placeholder for a binary industry classification such as GICS, BICS, NAICS, SIC, etc., where level = sector, industry, subindustry, etc. Multiple IndClass in the same alpha need not correspond to the same industry classification).

\subsection{Settings}
\noindent (1) The initial fundamental and technical indicators of stocks are transformed into 101 Alphas based on the formulas provided in "101 Formulaic Alphas" \cite{Kakushadze2015101FA}. These 101 alphas serve as the initial inputs $x$ for our model.

\noindent (2) Testing set refers to the data from the last 70 days, while the training set includes all data before that. Additionally, a validation set was created by randomly selecting 5\% of the training dataset. 

\noindent (3) The dimension $d$ of the model is 512. The multiplier $m$ is 4 and divisor k is 8.

\noindent (4) The dropout rate is 15\%.

\noindent (5) The output of total model is a one-dimension value, which is transformed by the last linear transformation: $\mathbb{R}^d\rightarrow\mathbb{R}$. The output of 1 means the expected excess return of stocks is positive and 0 means negative.

\subsection{Results}
We benchmark our model against Linear regression, Simple MLP, Stack MLP and Deep MLP. Here Simple MLP is the Multilayer Perceptron with single hidden layer, Stack MLP is stacking the simple MLP modules over five times, and Deep MLP is a variation of the Stack MLP module empowered by batch normalization, skip connection, and the dropout strategies. 

\begin{figure}[h]
%\vspace{-0.3cm}
\centering
\hspace{0.01cm}
\subfigure[Simple MLP]{
\begin{minipage}[t]{0.48\linewidth}
\centering
\includegraphics[scale=0.25]{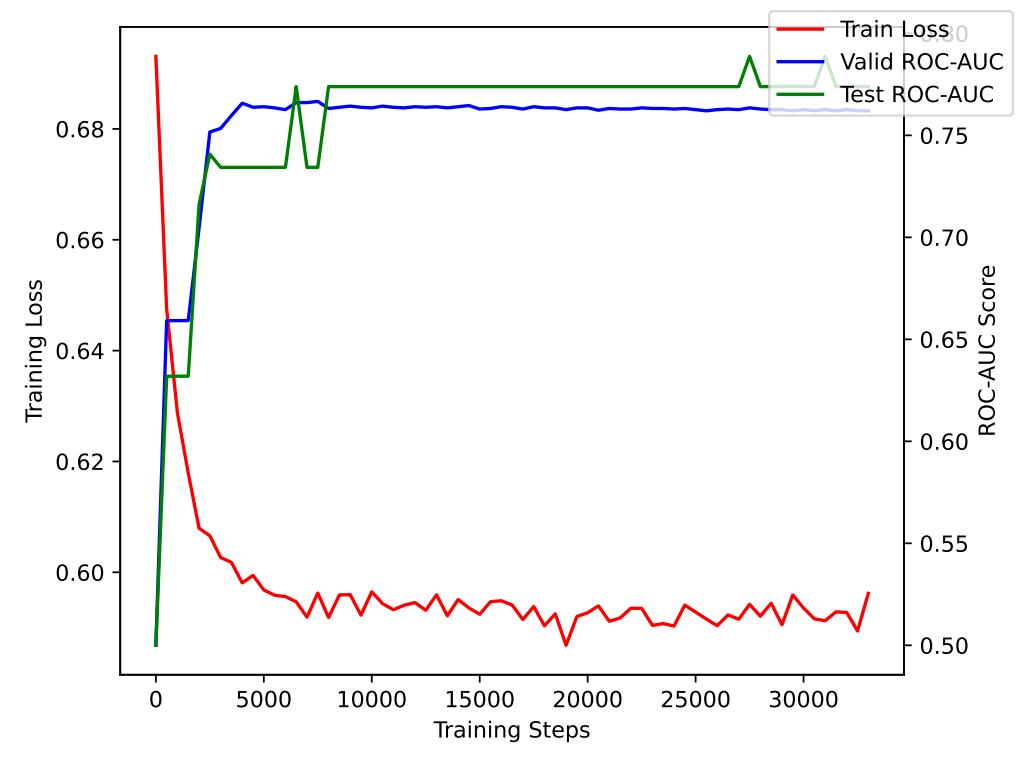}\label{val}
%\caption{fig}
\end{minipage}%
\hfill
}%
\hspace{-0.1cm}
\subfigure[Stack MLP]{
\begin{minipage}[t]{0.48\linewidth}
\centering
\includegraphics[scale=0.25]{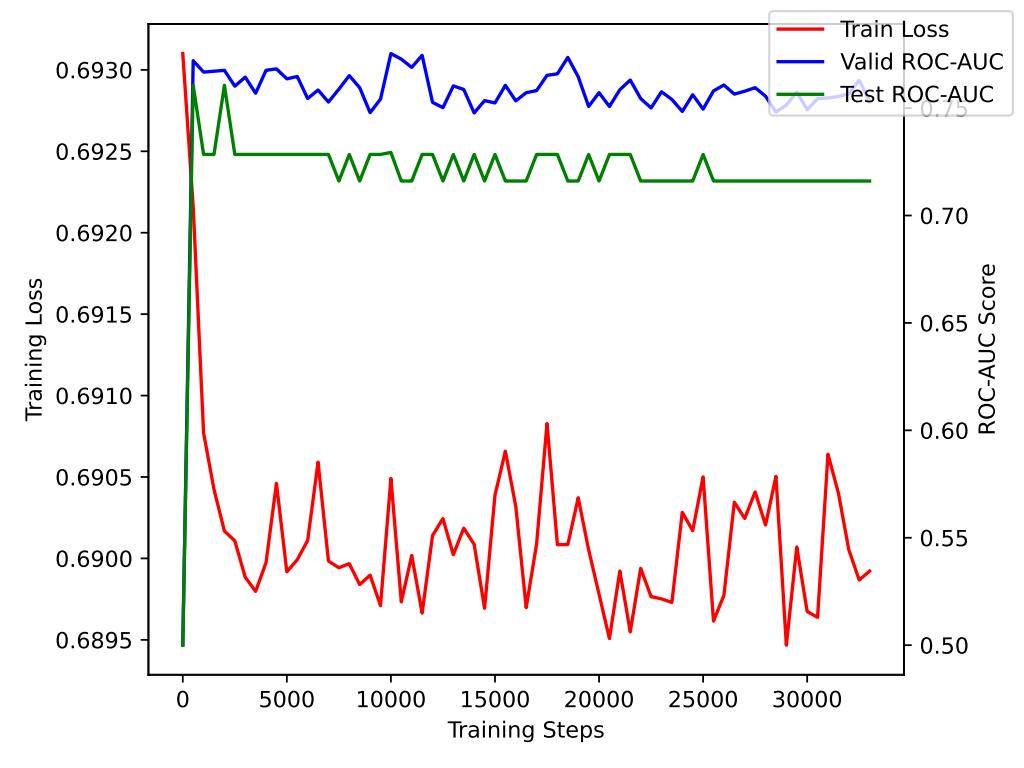}\label{val}
%\caption{fig}
\end{minipage}%
\hfill
}\hspace{0.1cm}

\subfigure[Deep MLP]{
\begin{minipage}[t]{0.48\linewidth}
\centering
\includegraphics[scale=0.25]{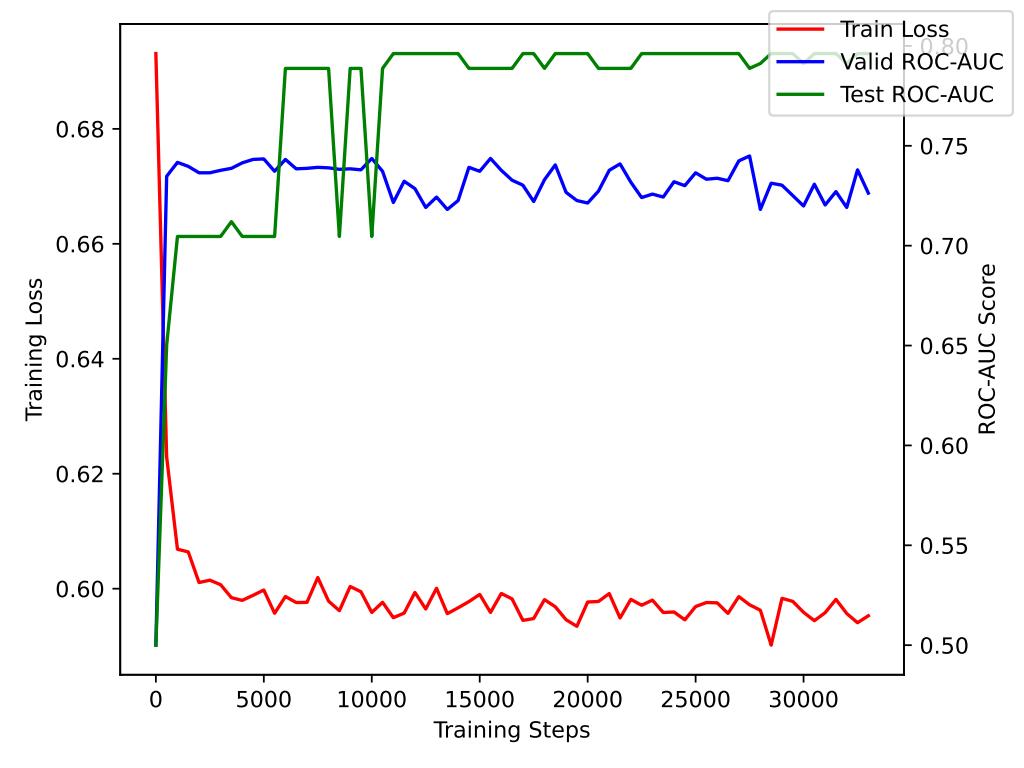}\label{val}
%\caption{fig}
\end{minipage}%
\hfill
}%
\subfigure[GatedDeepMLP]{
\begin{minipage}[t]{0.48\linewidth}
\centering
\includegraphics[scale=0.25]{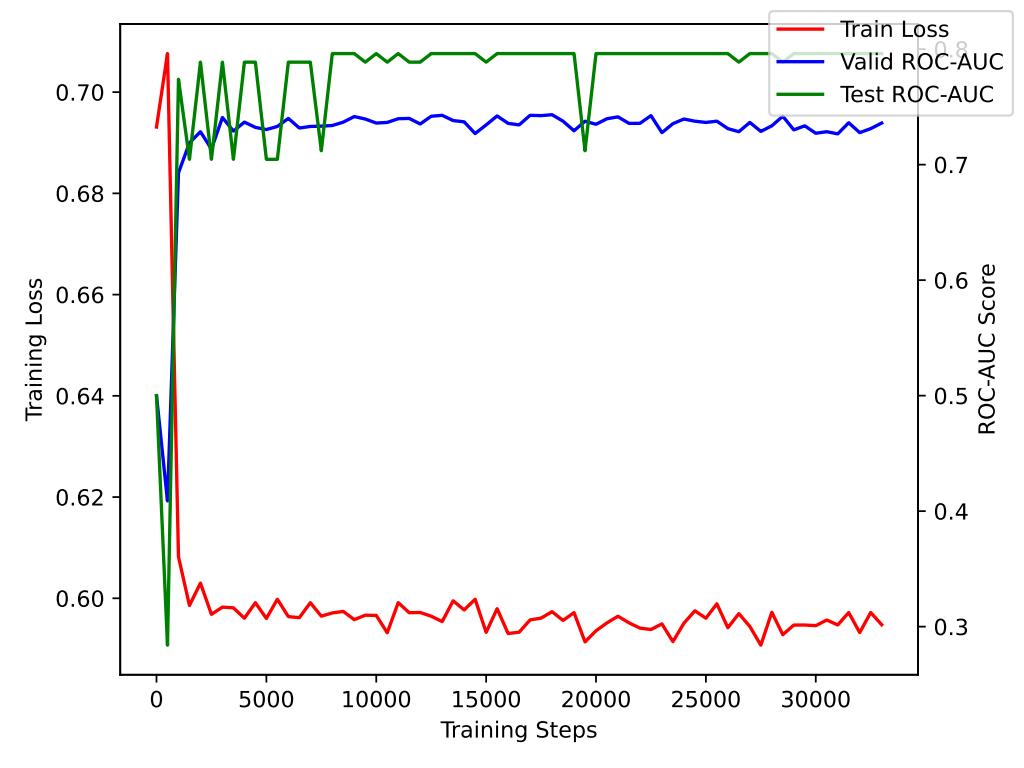}\label{val}
%\caption{fig}
\end{minipage}%
\hfill
}%

\centering
\caption{Our data is partitioned into training, testing and validation sets. The model is validated every 250 training steps using the validation set. After 3 epochs of training, the test result is reported by the model weight with the best validation performance.}
\end{figure}

\begin{table}[!ht]
\label{compare}
\centering
\caption{The ROC-AUC scores of four models.}
\setlength{\tabcolsep}{4mm}{
\begin{tabular}{ccc}

\hline
Model         & Valid & Test \\ 
\hline
Linear        & 0.7681        & 0.6319       \\
Simple MLP    & 0.7620        & 0.7343       \\
Stack MLP     & 0.7531        & 0.7294       \\
Deep MLP      & 0.7263        & 0.7887       \\
GatedDeep MLP & 0.7360        & 0.7962       \\ 
\hline
\end{tabular}}
\end{table}

\paragraph{Simply stacking layers could lead to degradation problem.} In Table\ref{compare}, the ROC-AUC score for the testing set of Stack MLP is 0.7294, which is lower than Simple MLP (0.7343). It shows that simply adding hidden layers is invalid for improving the model's learning ability. The Stack MLP model's AUC-ROC after 25000 steps did not improve significantly compared to its initial value.
\paragraph{Deeper models is beneficial in predicting.} Both Deep MLP and GatedDeep MLP perform better than Simple MLP and achieve high ROC-AUC scores for the testing set, which are 0.7887 and 0.7962 respectively.
\paragraph{Batch Normalization, Residual Connection and Dropout could enhance the model's robustness.} Deep MLP significantly outperforms Stack MLP. Besides higher ROC-AUC score, it is shown in Figure 6 that the training loss of Deep MLP obviously diverges and its ROC-AUC scores for the testing set quickly improves after 5000 steps.
\paragraph{The gated activation layer can improve the performance of model.} GatedDeepMLP outperforms the other three models on predicting the future excess return of stocks. Its ROC-AUC score (0.7962) is 1\% higher than Deep MLP.

\section{Future Works}

%IC即信息系数（Information Coefficient），表示所选股票的因子值与股票下期收益率的截面相关系数，通过 IC 值可以判断因子值对下期收益率的预测能力。信息系数的绝对值越大，该因子越有效。IC为负表示因子值越小越好，IC为正表示因子值越大越好。
%cosine similarity
\noindent The IC value is a metric that reflects the predictive power of a factor value on future returns. A larger absolute value of IC indicates a more effective factor. A negative IC indicates that a smaller factor value is better, while a positive IC indicates that a larger factor value is better.\\

\noindent Inspired by the concept of Information Coefficient (IC) which measures the correlation between a factor value and the cross-sectional next-period returns of a set of stocks, we have developed a new loss function: 
\begin{equation}
\begin{aligned}
L =  & \min\sum_{m=1}^M\sum_{t=1}^T \Bigg(\sum_{i=1}^N Sc(c_{i,m,t}, ret_{m,t+\Delta t}) \\
     & -\sum_{i=1,j=1,i\neq j}^N Sc(c_{i,m,t}, c_{j,m,t})\Bigg) \\
\end{aligned}
\end{equation}
%\begin{equation}
%    L=\mathop{\min}\sum_{m=1}^M\sum_{t=1}^T(\sum_{i=1}^N Sc(c_{i,m,t}, ret_{m,t+\Delta t})-{\sum_{i=1,\atop %j=1,\atop i\neq j}^N Sc(c_{i,m,t}, c_{j,m,t})）)}
%\end{equation}
and
\begin{equation}
    Sc(A,B) = \frac{A \cdot B}{||A|| ||B||}
\end{equation}
where:
\begin{itemize}
\item[] ${c_{i,m,t}}$ is the value of factor $i$ for stock $m$ on day $t$;
\item[] $\{ret_{t+\Delta t,m}\}$ is the excess return of stock $m$ on day $t+\Delta t$;
\item[] ${Sc(c_{i,m,t}, ret_{m,t+\Delta t})}$ is the cosine similarity between the value of factor $i$ on day $t$ and the future excess return of stock $m$ on day $t+\Delta t$;
\item[] ${ Sc(c_{i,m,t}, c_{j,m,t})}$ is the cosine similarity between the the value of factor $i$ on day $t$ and the value of factor $j$ on day $t$. It is designed to address the issue of high correlation between the chosen factors. When factors are highly correlated, it can lead to problems such as multicollinearity and overfitting, which can affect the stability and reliability of the model.
\end{itemize}

\noindent Due to the high complexity of the proposed loss function, currently available computational resources are insufficient to run the experiments. However, future upgrades in computational power may enable the execution of this experiment.

%%%%%%%%%%%%%%%%%%%%%%%%%%%%%%%%%%%%%%%%%%%%%%%%%%%%%%%%%%%%%%%%%%%%%%%%%%%%%%%%%%%%%%%%%%%%%%%%%%%%%%%%%%%%%%%%
\section{Conclusion}
In this study, we focus on the research problem about forecasting the trends of the asset's excess returns. Specifically, we propose a deep neural network with gated activation layer to combine the manually designed factors. Our experimental results prove that the proposed Gated Deeper MLP model significantly outperform the Linear Regression and naive two-layers MLP model. We also prove the effective of the new deep learning strategies in enhancing the training robustness and alleviating the overfitting problem. We believe our findings from this paper could inspire further work in this challenging problem.

\bibliography{2_custom}
\bibliographystyle{acl_natbib}

\end{document}